# Machine-Learned Atomic Cluster Expansion Potentials for Fast and Quantum-Accurate Thermal Simulations of Wurtzite AlN

Guang Yang[1,†], Yuan-Bin Liu[2,†], Lei Yang[1], Bing-Yang Cao[1,*]

[1] *Key Laboratory for Thermal Science and Power Engineering of Ministry of Education, Department of Engineering Mechanics, Tsinghua University, Beijing 100084, China*

[2] *Inorganic Chemistry Laboratory, Department of Chemistry, University of Oxford, Oxford OX1 3QR, UK*

**Abstract:**

Thermal transport in wurtzite aluminum nitride ($w$-AlN) significantly affects the performance and reliability of corresponding electronic devices, particularly when lattice strains inevitably impact the thermal properties of $w$-AlN in practical applications. To accurately model the thermal properties of $w$-AlN with high efficiency, we develop a machine learning interatomic potential based on the atomic cluster expansion (ACE) framework. The predictive power of the ACE potential against density functional theory (DFT) is demonstrated across a broad range of properties of $w$-AlN, including ground-state lattice parameters, specific heat capacity, coefficients of thermal expansion, bulk modulus, and harmonic phonon dispersions. Validation of lattice thermal conductivity is further carried out by comparing the ACE-predicted values to the DFT calculations and experiments, exhibiting the overall capability of our ACE potential in sufficiently describing anharmonic phonon interactions. As a practical application, we perform a lattice dynamics analysis using the potential to unravel the effects of biaxial strains on thermal conductivity and phonon properties of $w$-AlN, which is identified as a significant tuning factor for near-junction thermal design of $w$-AlN-based electronics.

**Keywords:** Wurtzite aluminum nitride, thermal conductivity, atomic cluster expansion, machine learning interatomic potential, strain engineering

[†] These authors contributed equally to this work.

[*] Corresponding author: caoby@mail.tsinghua.edu.cn

# 1. Introduction

Wurtzite aluminum nitride (*w*-AlN) emerges as a promising semiconductor, distinguished by various exceptional characteristics. These include an ultrawide bandgap[1–3] (~ 6.1 eV), a large critical electric field[1,3] (~ 15 MV/cm), a high sound velocity[4] (~ 11 km/s), large piezoelectric coefficients[5], relatively high thermal conductivity[6,7] ($\kappa$ ~ 300 W/m K), and a lattice similar with other semiconductors such as *w*-GaN. The large critical electric field stemming from the ultrawide bandgap results in Baliga's figure of merit (FOM) and Johnson's FOM of *w*-AlN significantly surpassing those of *w*-GaN or *β*-Ga$_2$O$_3$[1,3], thereby establishing *w*-AlN as a remarkable candidate for novel high-power or radio frequency (RF) electronics[8,9]. Meanwhile, the ultrawide bandgap of *w*-AlN facilitates developments in deep-ultraviolet photonics[3,8,10,11]. The high sound velocity and piezoelectric performance render *w*-AlN suitable for fabricating micro-electromechanical system (MEMS) based resonators and filters, which are extensively applied in 5G communications[12–15]. In addition, the thermal conductivity of *w*-AlN is considered satisfactory, hence *w*-AlN sometimes serves as high-$\kappa$ substrates for high-power devices to improve the heat dissipation performance[9,16,17]. As depicted in Figure 1, the bandgap and thermal conductivity of *w*-AlN are compared to other representative materials[2], further exhibiting the significant promise of *w*-AlN.

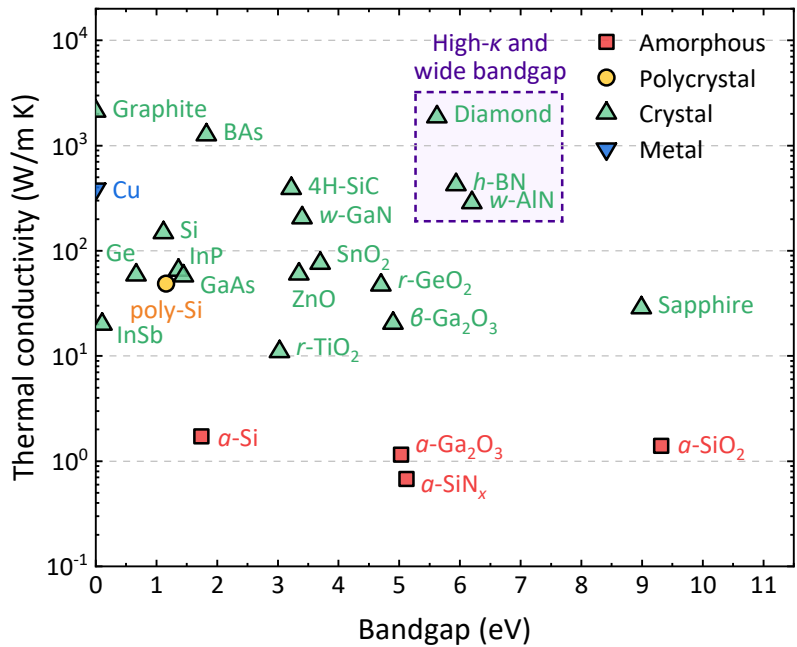

**Figure 1.** Room temperature thermal conductivities of different materials vs. their electronic bandgaps, including amorphous materials (e.g., *a*-Si, *a*-Ga$_2$O$_3$), polycrystals (poly-Si), metals (Cu), and nonmetallic crystals (e.g., *w*-AlN, *w*-GaN). The data are referred to literature[2,3,18–21]. This plot reveals that *w*-AlN lies in the range of promising comprehensive performance (high thermal conductivity and ultrawide bandgap). Note that the average thermal conductivities are chosen for anisotropic materials (e.g., $\beta$-Ga$_2$O$_3$, $r$-TiO$_2$).

Though the inherent characteristics of $w$-AlN are excellent, further studies on its properties remains necessary for developing next-generation electronics, especially in the aspects of: (i) clarifying how the complex environment affects the physical properties, (ii) tuning the physical properties on demand, and (iii) analyzing the growth processes. One specific issue of interest pertains to the effects of lattice strains on the thermal properties of $w$-AlN. Since $w$-AlN is extensively used as the nucleation layer within the GaN high electron mobility transistors (HEMTs) to buffer lattice mismatch between $w$-GaN epilayers and substrates[22–25], the residual stress within AlN is inevitable. However, the correlations between lattice strains with the thermal conductivity and phonon bands of $w$-AlN remain vague, which may affect the heat dissipation and reliability of corresponding devices[26,27].

In addition to experimental approaches, atomistic simulations act as another avenue for gaining insights into the physical properties of novel materials[28], which is traditionally represented by two techniques[29], i.e., the first-principle calculations based on density functional theory (DFT) and the molecular dynamics (MD) simulations based on empirical potentials. Nevertheless, high computational cost limits the DFT methods for modeling transport properties, while the MD simulations based on simple empirical potentials are less accurate than DFT[30]. For $w$-AlN, several empirical potentials have been proposed, including the Stillinger-Weber (S-W)[31], Tersoff[32], Vashishta[33], and COMB3[34] models, etc. However, each empirical potential generates divergent lattice parameters or phonon dispersions from those of DFT[33–35]. When predicting thermal conductivity, it is fundamental to accurately describe both harmonic and anharmonic interactions of phonons[36–38]. This places a heightened demand on the accuracy of interatomic potentials for $w$-AlN.

In recent years, machine learning (ML) interatomic potentials have attracted significant attention by effectively balancing computational efficiency with accuracy. A wealth of literature has shown a well-built ML potential trained with the DFT reference data can provide an unbiased representation of potential energy surfaces and simultaneously exhibit strong transferability[36,39,40]. More importantly, the linear behavior in the computational cost of ML potentials enable them much higher efficiency and scalability than DFT methods. Until now, several ML potential models have been proposed, such as the neural network potential (NNP)[41,42], Gaussian approximation potential (GAP)[40,43,44], spectral neighbor analysis potential (SNAP)[45,46], deep potential (DP)[47,48], moment tensor potential (MTP)[49,50], atomic cluster expansion (ACE) potential[51,52], neural equivariant interatomic potential (NequIP)[53], Allegro[54], and MACE[55,56], etc. The ACE potential is one of the most computationally efficient and quantum-accurate models available[52], and is also suitable for performing large-scale simulations on CPU platforms. Hence, the ACE is chosen in this work.

Here, we introduce a machine-learned ACE potential[51,52,57] for *w*-AlN, aiming to facilitate the atomistic simulations of its thermal properties and gain insights into how tuning factors (such as lattice strains) influence the phonon pictures. The remainder of this paper is organized as follows. In Section 2, we concisely introduce the ACE methodology, and the construction of training database for *w*-AlN. In Section 3, we comprehensively demonstrate the accuracy of our ACE potential in predicting various thermal and mechanical properties of *w*-AlN by comparison with either DFT calculations or experiments. Then, our ACE potential is applied to unravel the correlations between thermal conductivities and biaxial strains of *w*-AlN. Essential conclusions of this study are presented in Section 4.

## 2. Methods

### 2.1 Atomic cluster expansion framework

In line with other common ML potentials, the ACE model also expresses the total energy of a given system as the sum of site energies,

$$E = \sum_i \varepsilon_i, \qquad (1)$$

in which each $\varepsilon_i$ depends on its local atomic environment within a given cutoff radius $r_{\text{cut}}$. Different from other many two-, three-, and many-body descriptors that are not strictly complete, the ACE framework provides an efficient representation of local atomic environments by means of a complete linear basis of body-ordered symmetric polynomials[51,52,57].

Specifically, atomic energy contribution $\varepsilon_i$ in the ACE model is represented as

$$\varepsilon_i = F\left(\varphi_i^{(1)}, \ldots, \varphi_i^{(p)}\right), \qquad (2)$$

where $F$ is a generalized nonlinear function to be supplied, and $\varphi_i^{(p)}$ is the fundamental building block of ACE which is expanded by body-ordered functions within the set of neighbors for each atom $i$:

$$\varphi_i^{(p)} = \sum_{\mathbf{znlm}} \tilde{\mathbf{c}}_{z_i\mathbf{znlm}}^{(p)} \mathbf{A}_{i\mathbf{znlm}}. \qquad (3)$$

The $\tilde{\mathbf{c}}_{z_i\mathbf{znlm}}^{(p)}$ denotes expansion coefficients, and vectors $\mathbf{z}$, $\mathbf{n}$, $\mathbf{l}$, and $\mathbf{m}$ contain atomic species, indices for radial functions, and indices for spherical harmonics, respectively. The permutation-invariant many-body basis functions $\mathbf{A}_{i\mathbf{znlm}}$ are represented as

$$\mathbf{A}_{i\mathbf{znlm}} = \prod_{t=1}^{v} A_{iz_t n_t l_t m_t}, \qquad (4)$$

where the order of the product $v$ determines the body order of a basis function. Meanwhile, the atomic base $A_{iz_t n_t l_t m_t}$ is given as

$$A_{iz_t n_t l_t m_t} = \sum_j \delta_{zz_j} \phi_{z_i z_j nlm}(\mathbf{r}_{ij}), \qquad (5)$$

in which $\mathbf{r}_{ij}$ is the relative position of neighbor atoms, and the one-particle basis $\phi_{z_i z_j nlm}$ consists of spherical harmonics functions $g_{nl}^{z_i z_j}$ and radial functions $Y_{lm}^{z_i z_j}$:

$$\phi_{z_i z_j nlm} = g_{nl}^{z_i z_j} Y_{lm}^{z_i z_j}. \qquad (6)$$

It is noting that the expansion coefficients $\tilde{c}^{(p)}_{z_i znlm}$ in Eq. (3) cannot be directly used for model fitting because the many-body basis functions **A** do not satisfy rotational symmetries. By utilizing generalized Clebsch–Gordan coefficients to couple the elements of the basis function **A**, an invariant basis function **B** is obtained, **B** = **CA**. Consequently, a linear model invariant to translation, rotation, and permutation of like atoms, can be written for the site energy of ACE:

$$\varphi_i = \mathbf{c}^T \mathbf{B}. \tag{7}$$

The coefficients **c** are free model parameters that can be optimized during fitting. Further information about the ACE architecture can be found in the literature[51,52,57]. In this work, we employ the software package Pacemaker[58] for the parametrization of ACE potential. The final ACE model associated with the detailed hyperparameters is freely available in the Supplementary Material.

## 2.2 Construction of the training database

Structures in the training database are obtained from MD trajectories. Here, the empirical S-W potential[31] is adopted to carry out the MD simulations. This approach allows for sampling over extended time scales (~ ns) to ensure structural diversity, while bypasses the computationally intensive *ab initio* MD and makes the sampling more efficient. In detail, the initial structure consists of a 3 × 3 × 3 supercell containing 108 atoms. The cell is then expanded or compressed with a scaling factor ranging from 0.95 to 1.05 on each lattice constant. Next, a series of MD simulations for each cell are performed with the canonical (NVT) ensemble at temperatures of 100 K, 500 K, and 1000 K, respectively. For each temperature, a 1-ns trajectory is produced, from which structures are sampled at uncorrelated intervals of 150 ps. Finally, a total of 13,608 local atomic environments are collected from all 21 MD trajectories.

All generated structures are subsequently subjected to single-point DFT calculations to obtain well-converged reference energies and forces for training. The DFT calculations are performed with the Vienna *ab initio* simulation package (VASP)[59]. Exchange and correlation are treated by using the PBEsol functional[60] with a projector augmented wave method[61]. Moreover, we adopt Gaussian smearing of 0.05 eV width to electronic levels, a 600-eV cutoff for plane wave expansions, and a maximum spacing of 0.2 Å$^{-1}$ for meshing the reciprocal space. The total energy is attained with a convergence criterion of less than $10^{-6}$ eV in the self-consistent electronic iterations.

## 3. Results and Discussion

### 3.1 Performance of the ACE potential for wurtzite AlN

The accuracy of our ACE potential is validated through a comparison with DFT-predicted energies and forces. The training and testing datasets comprise 36,612 and 4,212 atomistic force components, respectively. In Figure 2, we present a comparison of the total energies and atomic forces predicted by our potential with those derived from DFT. Notably, our ACE potential effectively reproduces the total energies, showcasing a remarkably low root-mean-squared error (RMSE) of 0.13 meV/atom for the testing datasets. Furthermore, the interatomic forces in the testing datasets are accurately predicted, with a relatively low RMSE of 5.01 meV/Å. The accuracy of both energy and force predictions achieved with our ACE potential is comparable to those of other reported works[40,43,52,62,63], which demonstrates that our ACE potential serves as a robust representative of the DFT potential energy surface.

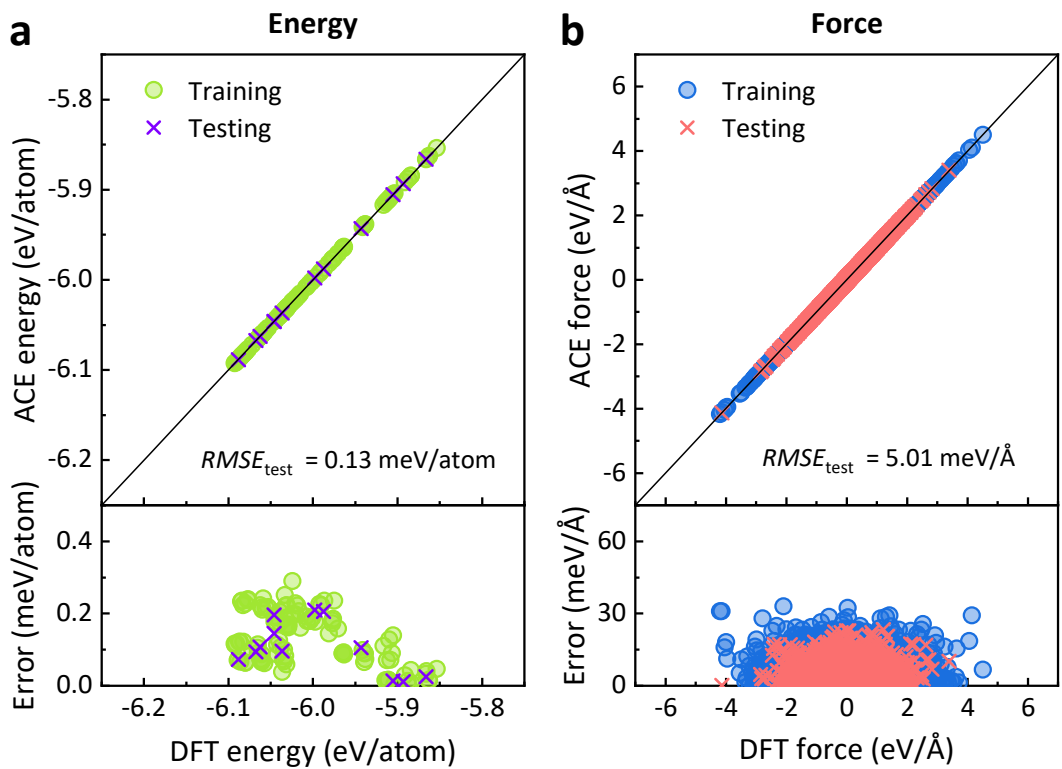

**Figure 2.** Comparison of DFT-computed and ACE-predicted (a) total energies and (b) interatomic forces for $w$-AlN. Here, "Error" represents the absolute error.

Then, we assess the capability of our ACE potential to model the physical properties of $w$-AlN. Our evaluation begins by employing the ACE to predict lattice parameters, a fundamental property that significantly influences various intrinsic characteristics of a material. The lattice parameters produced by our ACE potential are as follows: $a = b = 3.11558$ Å, $c = 4.98513$ Å, and $\gamma = 120°$. These values are in excellent agreement with both our DFT calculations and experimental results[64] (Table 1). For comparison, lattice parameters are also predicted using the S-W potential[31] here as well as using the Tersoff potential elsewhere[35]. However, it's noteworthy that the relative errors in these predictions are considerably larger than those obtained with ACE.

| Method | Lattice parameter | | | Max relative error against Exp. |
|---|---|---|---|---|
| | $a/b$ (Å) | $c$ (Å) | $\gamma$ (°) | |
| ACE potential relaxation | 3.11558 | 4.98153 | 120 | 0.116% |
| SW potential[31] relaxation | 3.08002 | 5.02965 | 120 | 1.027% |
| DFT relaxation (PBEsol) | 3.11288 | 4.98247 | 120 | 0.032% |
| Experiment[64] | 3.11197 | 4.98089 | 120 | / |

Table 1. Comparison of the lattice parameters of *w*-AlN determined by different methods.

The ACE potential is subsequently employed to predict the volumetric specific heat capacity ($C_V$), coefficients of thermal expansion, and the bulk modulus of *w*-AlN. $C_V$ is calculated using the Phonopy package[65–67]. Figure 3(a) summarizes the $C_V$ obtained from our ACE potential, the DFT calculations, and the experimental measurements[68]. Obviously, the data generated by our ACE potential exhibit strong agreement with both DFT and experiments.

Thermal expansion of *w*-AlN holds significant importance in practical applications, particularly when *w*-AlN is utilized as transition layers in III-V electronics, such as GaN HEMTs[22,23,25]. This property determines heteroepitaxial strains arising from the thermal mismatches during growth processes[22]. The coefficient of thermal expansion (CTE, denoted by $\alpha_E$) is defined as

$$\alpha_E = \frac{1}{V_T}\frac{\partial V_T}{\partial T}, \tag{8}$$

where $V$ denotes the volume of AlN unit cell, and $T$ represents temperature. The CTEs are calculated under the quasi-harmonic approximation[65] (QHA) implemented with Phonopy. As shown in Figure 3(b), the ACE potential quantitatively reproduces the thermal expansion determined by DFT calculations. Both ACE and DFT predict that the CTEs are strongly dependent on temperatures between 0 K and 1000 K. It is evident that as the temperature increases, the CTEs exhibit rapid growth up to ~ 300 K, after which the slopes decrease at higher temperatures.

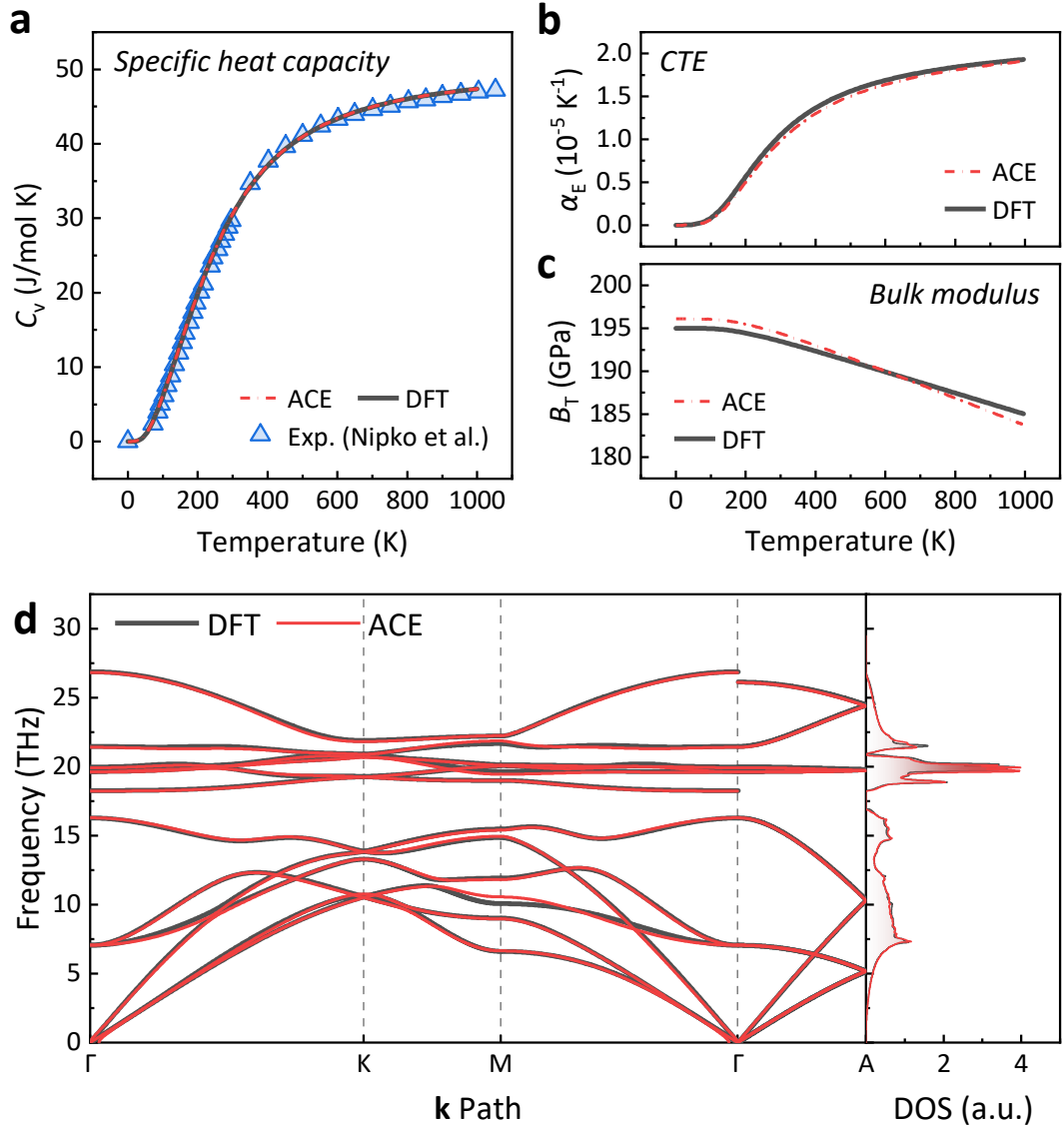

**Figure 3.** Temperature-dependent (a) specific heat capacity, (b) thermal expansion coefficients, and (c) bulk modulus of *w*-AlN. (d) Phonon dispersion and phonon density of states (DOS) of *w*-AlN at 0 K predicted by DFT and the ACE potential.

Moreover, a mechanical property, namely the bulk modulus ($B_T$) of *w*-AlN is also determined by ACE potential. As shown in Figure 3(c), the ACE potential quantitatively reproduces the $B_T$ as DFT calculations. Both ACE and DFT predict a subtle decrease in the modulus as the temperature increases from 0 K to 1000 K. In summary, it is obvious that our ACE potential is capable of accurately describing both thermal and mechanical properties.

The prediction of the phonon dispersions of a material is another crucial metric for the quality of a potential to describe lattice dynamics. We first calculate the second-order harmonic and third-order anharmonic interatomic force constants (IFCs) through the finite displacement method[69]. By combining the Phonopy package with the second-order IFCs calculated from our ACE and DFT, the phonon dispersions of $w$-AlN at 0 K are determined, as illustrated in Figure 3(d). As a characteristic of the ionic crystal, the splitting of LO–TO phonons at the Γ-point is observed, which is attributed to the long-range Coulomb interactions[36]. Since our ACE model does not encompass the Coulomb interactions, we have incorporated the non-analytical correction[26] into dynamical matrix to resolve the splitting of LO–TO phonons at the Γ point. Our results show that the ACE model accurately predicts the phonon frequencies at almost of all high-symmetry points and accurately captures the dispersion behavior of each phonon branch. Meanwhile, the phonon density of states (DOS) calculated by ACE is almost identical to the DFT results.

In fact, although the absence of Coulomb terms introduces some errors in force calculations, it will be shown later that the accuracy of our potential remains sufficiently high to yield thermal conductivity predictions. If the long-range Coulomb interactions become crucial for other calculations, we will properly incorporate the Coulomb interactions with fixed partial charges before fitting the ACE model. This adjustment will automatically resolve the LO–TO splitting issue, and improve the accuracy of force calculations[36]. Similar treatments for GaN can be found in the literature[43].

## 3.2 Thermal conductivity of wurtzite AlN

Herein, we utilize the second- and third-order IFCs to predict of the thermal conductivity ($\kappa$) of $w$-AlN. The Wigner transport equation (WTE)[70–72] is solved by the direct solution in Phono3py[66,73,74] package, with a 25×25×25 mesh for sampling the first Brillouin zone over temperatures within 100 – 600 K. Due to its wurtzite lattice, it is expected that the thermal conductivities of $w$-AlN exhibit approximate isotropy along both in-plane and cross-plane directions[27].

We also conduct ShengBTE[75] calculations including the four-phonon processes (4ph)[76] for comparison with the WTE. Since the computational cost is extremely huge with 4ph involved, we resort to the sampling-accelerated method[77] with the settings "num_sample_process_4ph = 1E5" and "num_sample_process_4ph_phase_space = 1E5". Meanwhile, a 16×16×16 mesh of the first Brillouin zone is set, and the value of "scalebroad" is taken as 0.1. The cutoff distances of the third- and fourth-order IFCs are selected to be the seventh nearest and the second nearest atom respectively. These mentioned parameters are large enough to make it converge to the rigorous calculations[75–78], though resulting in a negligible uncertainty due to random sampling.

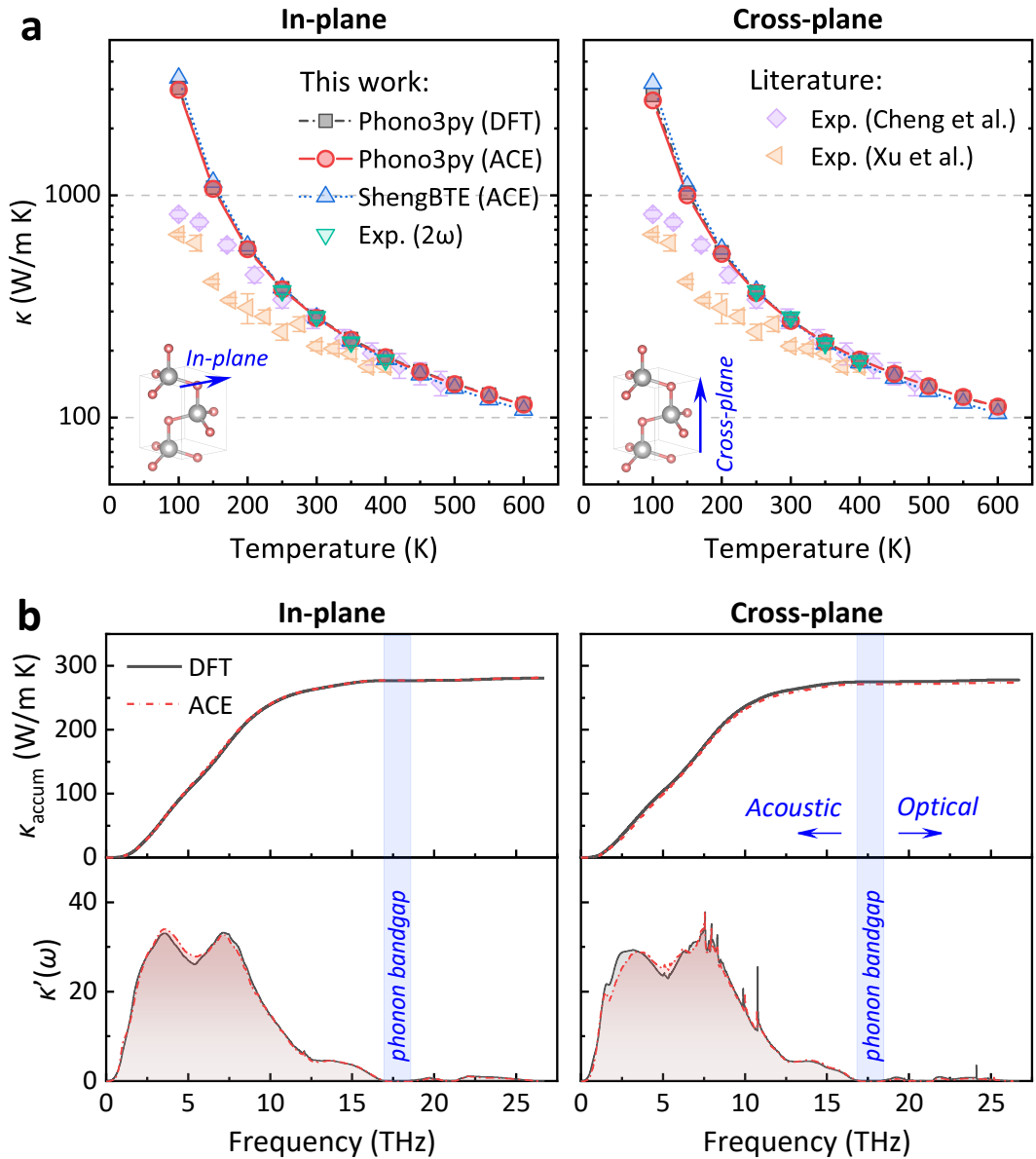

**Figure 4.** (a) Comparison of *w*-AlN's thermal conductivities along the in-plane and cross-plane directions between the ACE model, DFT calculations and experiments. Note that all the Phono3py calculations here have enabled the Wigner transport theory[70–72], while the ShengBTE calculations have included four-phonon effects[76].

The literature experiments[2,79] depicted here are all based on an isotropic assumption, while experiments based on our three-sensor 2ω method[7] are capable of deriving the anisotropic thermal conductivity directly. (b) Comparison of accumulated thermal conductivity of $w$-AlN between the ACE model and DFT calculations, as a function of phonon frequency at 300 K along the in-plane and cross-plane directions, calculated by Phono3py.

As illustrated in Figure 4(a), the ACE potential calculates the thermal conductivities of $w$-AlN along both in-plane and cross-plane directions. Though the physical pictures behind WTE and 4ph-scattering are different, their predictions are close indeed. The thermal conductivity predicted by ACE shows overall good agreements with our DFT calculations and experimental measurements (based on the three-sensor 2ω method proposed in our recent works[7,22,80]). Besides, the ACE-predicted values approximate the literature experiments[2,79] above room temperature as well, though the ACE overestimate thermal conductivities beneath 250 K owing to neglecting the phonon-defect scattering. Several theoretical and experimental studies[27,79,81,82] have discussed on this issue, researchers generally attribute the discrepancy between theoretical predictions and cryogenic experiments to the impacts of phonon-defects (e.g., point defects, dislocations, grain boundaries) scattering. Since the phonon-defects scattering rates are independent with temperature in principle[27], they will dominate the phonon transport at low temperatures where the anharmonicity (normal and Umklapp processes) is weak.

The literature experiments depicted here are all based on an isotropic assumption, whereas our three-sensor 2ω method enables the direct derivation of thermal conductivities along different directions[7]. The brief introduction on the three-sensor 2ω method can be found in the Supplementary Material.

Furthermore, we conduct an equilibrium molecular dynamics (EMD) simulation based on the ACE potential to predict the $w$-AlN thermal conductivity. The large quantum effects at low temperatures significantly influence the MD-calculated thermal conductivity[28,81,83–86], since the Debye temperature of $w$-AlN is quite high ($\Theta_D \sim 1000$ K)[87]. Hence, we only calculate the thermal conductivity at 1100 K, where the quantum effects could be reasonably neglected[84]. A bulk $w$-AlN system containing 4000 atoms is set as the initial configuration for EMD simulations conducted in the LAMMPS[88]. The length of each crystallographic direction reaching 10 unit cells, which achieves convergence for calculating the thermal conductivity of $w$-AlN[89,90].

Periodic boundary conditions are applied to all three directions to mimic the infinite size of structures. A total of 4.5 ns EMD simulations based on the trained ACE potential are performed with a time step of 1 fs. After equilibrating the system in the isothermal-isobaric (NPT) ensemble for 500 ps, the system is switched to the microcanonical (NVE) ensemble for 4 ns to collect heat flux data. To mitigate the impacts of statistical uncertainties and errors, we conduct ten independent EMD simulations with different initial velocity distributions to obtain the averaged thermal conductivities at 1100 K. The thermal conductivity here is considered isotropic along the in-plane directions, so we average the thermal conductivity values along the $x$, $y$ directions as the final in-plane thermal conductivity. The EMD results are $\kappa_{in}^{MD} = 51.0 \pm 5.8$ W/m K and $\kappa_{cr}^{MD} = 48.4 \pm 6.3$ W/m K, which are comparable to those of BTE calculations, as discussed in the Supplementary Material. On the AMD EPYC™ 7452 CPU platform, an average calculation efficiency of the ACE-based EMD simulations reaches 0.26 ms/MD-step/atom when calculating on a single thread (using GCC 9.1.0 compiler and LAMMPS 23-Jun-2022-Update3), which manifests the capability of ACE potential for large-scale molecular dynamics[52].

More detailed phonon transport characteristics of *w*-AlN are also calculated from the ACE potential by Phono3py. The accumulated thermal conductivity as a function of phonon frequency at 300 K is presented in Figures 4(b), which further verifies the accuracy of the ACE potential comparing to the DFT results. Obviously, the thermal conductivity is primarily contributed by acoustic phonon branches with frequencies below the "phonon bandgap", whose contributions exceed 98.5% for both in- and cross-plane values. In addition, we also calculate the accumulated thermal conductivity as a function of phonon mean free path (MFP) based on the single-mode relaxation time approximation, as detailed in the Supplementary Material. The discussions on MFP imply that the size effect of *w*-AlN film's thermal conductivity is crucial for practical applications, which limits the heat dissipation performance of corresponding electronic devices[22,23,25,91,92].

### 3.3 Influence of biaxial strain on thermal properties of wurtzite AlN

As discussed in Section 1, there is an inevitable residual strain (stress) inside *w*-AlN in its practical applications, especially when *w*-AlN serves as the transition layer for a GaN HEMT by forming the GaN/AlN/Substrate heterojunction. Owing to the mismatches in lattice and thermal expansion between three materials, residual strain and lattice defects inside the *w*-AlN are general significant. Therefore, based on the trained ACE potential of $w$-AlN, we proceed to study the strain effects on lattice thermal conductivity of $w$-AlN.

In practical applications of transition layers, two kinds of strain exist within $w$-AlN, i.e., in-plane biaxial strain perpendicular to the polarization axis, and cross-plane uniaxial strain along the polarization axis. Considering that in-plane stress is much more common than cross-plane stress in heterojunctions owing to the in-plane lattice mismatch[26], and the uniaxial strain hardly affects thermal conductivity[93], we only investigate the biaxial strain effects in this work. Here, the strain is applied by modifying the lattice constants of the structure, and then the structure is relaxed with in-plane lattice constant $a$ (or $b$) being settled. The biaxial strain is expressed by relative variation of the in-plane lattice constant,

$$\sigma_a = \frac{a - a_0}{a_0}. \tag{9}$$

Under in-plane biaxial strain states, the other lattice constant $c$ will also vary to achieve minimal stress, and finally an optimized structure in the cross-plane direction is formed. Also, the changes of crystal symmetries are not detected under strain states in this work, i.e., AlN maintains a wurtzite structure with the space group P6$_3$/mmc. The introduction of lattice strains and the structure optimization are performed using the atomic simulation environment (ASE)[94] package. Consequently, phonon properties as well as the thermal conductivity of $w$-AlN exhibit continuous changes under biaxial strain $\sigma_a$ states, ranging from −4% to 4%.

From a classical and intuitional perspective, compressive strain is expected to increase the thermal conductivity by augmenting the elasticity modulus and acoustic velocity[26]. The results, shown in Figure 5, reveal that both in- and cross-plane thermal conductivity of $w$-AlN decrease remarkably under the +4% biaxial strain state (tensile) at room temperature, with the average thermal conductivity decreasing by 40%. Conversely, under the −4% biaxial strain state (compressive), the average thermal conductivity increases by 30%. In the temperature range of 200 – 400 K, lattice thermal conductivity decreases significantly, however, the decrease/increase trend depending on biaxial strain is nearly constant at different temperatures (Figure 5(b)), implying that the influences of lattice strain and temperature on thermal conductivity should be independent. Meanwhile, the temperature dependence of thermal conductivity follows a similar trend under different biaxial strain states.

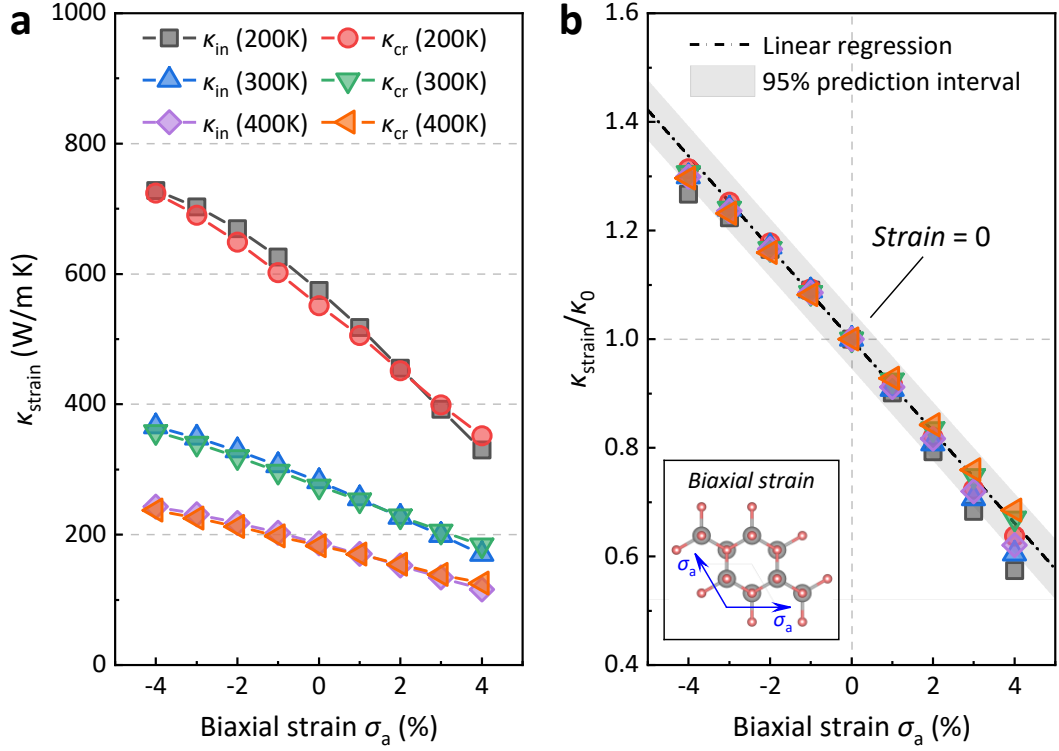

**Figure 5.** The effects of biaxial strains on thermal conductivities of *w*-AlN. (a) The dependence between thermal conductivities and the biaxial strains at 200 − 400 K, and (b) the variation of relative thermal conductivities at each temperature, i.e., the strained $\kappa_{\text{strain}}$ divided by the strain-less $\kappa_0$ at the same temperature.

Then, we investigate the changes of phonon properties under different strain states, to elucidate the correlation with their thermal conductivity. Based on the phonon BTE, lattice thermal conductivity can be expressed as[26]

$$\kappa_L^{\alpha\beta} = \sum_{\mathbf{q},\omega} C_{\mathbf{q},\omega} v_{\mathbf{q},\omega}^{\alpha} v_{\mathbf{q},\omega}^{\beta} \tau_{\mathbf{q},\omega}, \qquad (10)$$

in which **q**, $\omega$ denote the phonon branch and the frequency of a specific phonon mode, respectively. Thus, lattice thermal conductivity primarily depends on the three variables, namely volumetric specific heat $C_{\mathbf{q},\omega}$, group velocity $v_{\mathbf{q},\omega}^{\alpha}$ ($v_{\mathbf{q},\omega}^{\beta}$), and relaxation time $\tau_{\mathbf{q},\omega}$ of each phonon mode. Note that all these parameters are determined by the phonon dispersions, as illustrated in the Supplementary Material.

Figures 6(a)-(b) illustrate the phonon harmonic properties under different strain states, while Figures 6(c)-(d) depict the anharmonic features. Volumetric specific heat of phonon modes reflects the energy level of a crystal system, varying with the phonon dispersions and volume of unit cell

correspondingly. As shown in Figure 6(a), slight decreases of mode specific heat with tensile biaxial strains occur, implying a positive correlation to the change of thermal conductivity. In Figure 6(b), phonon group velocity gradually increases from the tensile to compressive strain states, which is consistent with the increased thermal conductivity. Figure 6(c) shows that phonon relaxation time decreases under the tensile strain state and increases under the compressive strain state, which is consistent with the thermal conductivity variations as well. According to the results of phonon DOSs in Supplementary Material, the phonon bandgaps are 0.65, 1.35, and 2.40 THz under tensile, free, and compressive strains, respectively. This determines the variations of relaxation time in principle, since a smaller (larger) phonon bandgap will enable more (less) available three-phonon scattering channels, thereby enhancing (suppressing) the three-phonon anharmonic scattering processes[26,95].

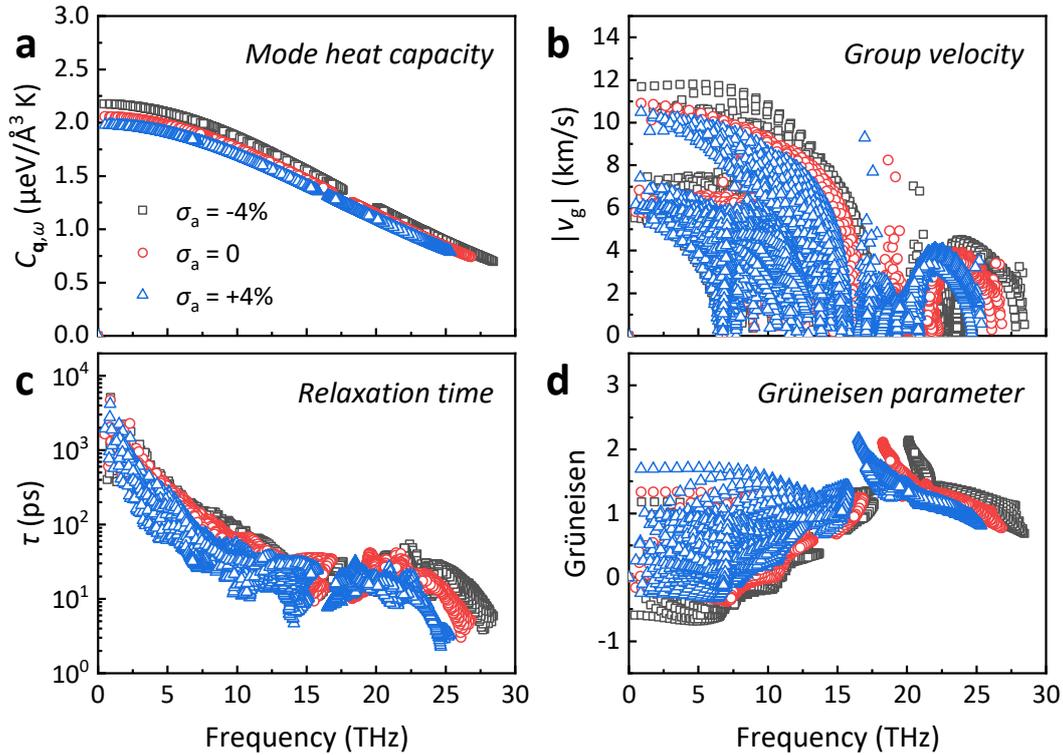

**Figure 6.** Phonon properties under different biaxial strain states. (a) Mode volumetric specific heat, (b) phonon group velocity, (c) phonon relaxation time, and (d) Grüneisen parameters.

The Grüneisen parameter, an indicator of lattice anharmonicity[26,27], is further analyzed here to back up the influences of strains on anharmonic scattering processes (Figure 6(d)). Obvious increases (decreases) in the Grüneisen parameters of acoustic phonon branches are observed under the tensile (compressive) strain state, in line with the variations of phonon bandgaps. More discussions on how and why lattice strains affect the phonon anharmonicity can be found in the Supplementary Material. Consequently, the consistent variations in mode specific heat, group velocity, relaxation time, and anharmonicity lead to significant changes in the lattice thermal conductivity[26] of *w*-AlN under different strain states.

The highlighted impacts of lattice strains on phonon properties should be favorable for tuning the heat dissipation performance of corresponding *w*-AlN-based electronic devices. By carefully engineering the residual strains within heteroepitaxial structures via annealing[96,97] or selecting substrates with specific lattice structures[26,98], it is viable to optimize the lattice thermal properties on demand, facilitating near-junction thermal optimization of semiconductor devices[91,99].

# 4. Conclusions

We have developed an ACE potential based on machine learning for atomistic simulations of monocrystalline $w$-AlN. Our ACE potential exhibits remarkable accuracy in reproducing the DFT potential energy surface of $w$-AlN, achieving an energy RMSE of ~0.13 meV/atom and a force RMSE of ~5.01 meV/Å for both Al and N atoms. Subsequently, the predictive power of ACE is demonstrated across a variety of properties of $w$-AlN, including ground-state lattice parameters, specific heat capacity, coefficients of thermal expansion, bulk modulus, phonon dispersions, as well as thermal conductivity. All these results show excellent agreement with the DFT calculations and experimental results, demonstrating that the ACE model sufficiently describe both harmonic and anharmonic phonon properties.

The lattice strain is proven as a significant tuning factor for thermal design of heteroepitaxial electronic devices. As a practical application of the ACE potential, we perform lattice dynamics simulations to unravel the effects of biaxial strains on thermal conductivity of $w$-AlN. The results indicate that a 4% biaxial tensile (compressive) strain approximately cause a 40% decrease (30% increase) in the thermal conductivity of $w$-AlN, while the influences of lattice strain and temperature on thermal conductivity appear to be independent. The investigations into phonon pictures under different strains reveal that the consistent variations in phonon mode heat capacity, group velocities and relaxation times predominantly contribute to the variation of thermal conductivity, while all these factors stem from the variations in phonon band structures. Thus, it is feasible to facilitate near-junction thermal optimization of devices via strain engineering on phonon bands. The findings here are beneficial for the development of next-generation electronic devices.

## Supplementary Material

See the Supplementary Material for brief introductions to the experiments, additional discussions on phonon properties, and the ACE potential files developed for simulations in both LAMMPS and ASE Calculators.

## Data availability

The ACE potential is attached in the Supplementary Material, which can be used with the ML-PACE package on https://github.com/ICAMS/lammps-user-pace. The data that support the findings of this study are available from the corresponding author upon reasonable request.

## Acknowledgments

This work was financially supported by the National Natural Science Foundation of China (Grant No. U20A20301, 52327809, 52250273, 51825601).